\documentclass[twocolumn]{aastex6}

\usepackage{amsmath}
\usepackage{epsfig}
\usepackage{graphicx}
\usepackage{subfigure}
\usepackage{url}
\usepackage{natbib}
\usepackage{float}
\usepackage{multirow}
\usepackage{cases}
\usepackage{color}

\begin{document}


\title{Is the Spiral Galaxy a Cosmic Hurricane?}


\author{CAO Zexin\altaffilmark{1}}
\affil{College of Science, Shenyang Aerospace University, Shenyang 110136, China}
\author{LIU Ling,   ZHENG Tingting}
\affil{College of Physics Science and Technology, Shenyang Normal University, ShenYang 110034, China}

\altaffiltext{1}{caozx@sau.edu.cn}

\begin{abstract}
It is discussed that the formation of the spiral galaxies is driven by the cosmic background rotation, not a result of an isolated evolution proposed by the density wave theory. To analyze the motions of the galaxies, a simple double particle galaxy model is considered and the Coriolis force formed by the rotational background is introduced. The numerical analysis shows that not only the trajectory of the particle is the spiral shape, but also the relationship between the velocity and the radius reveals both the existence of  spiral arm and the change of the arm number. In addition, the results of the three-dimensional simulation also give the warped structure of the spiral galaxies, and shows that the disc surface of the warped galaxy, like a spinning coins on the table,  exists a whole overturning movement. Through the analysis, it can be concluded that the background environment of the spiral galaxies have a large-scale rotation, and both the formation and  evolution of  hurricane-like spiral galaxies are driven by this background rotation.
\end{abstract}




\section{Introduction}\label{sect:intro}
The beautiful and unusual spiral structure of spiral galaxy attracts people's attention all the time. Conjecture and hypothesis of the formation mechanism have been made by many people, and a lot of investigations and verifications have been carried out. Among those works, the density wave theory developed by C.C. Lin and others has been acknowledged by many people(\citet{lin1964,lin1966}). The density wave theory holds that the rotation speed and density of the stars around the center vary in a wave motion, which forms a density wave(\citet{pasha2004}). 

The wave travels both along the center circle and at the same time along the radius, therefore, the peaks with the highest density are in a spiral shape and thus form a spiral arm. After coming into the arm area the star materials would slow down their velocity because of the dense stars and the strengthened gravitational field; on the other hand, the slowed down velocity of the stars would make the stars to crowd together, which would cause the density increases, and then the gravitational field would be strengthened further. Thus the stability of the arms which is made up of areas of greater density would be maintained self-consistently. According to the model of the density wave theory, the quasi-steady solutions can be obtained from the gravitational field equations and dynamical equation of galaxies, to demonstrate the spiral structure of the galaxies.

The spiral structure observed from the spiral disc galaxies, such as the galaxy M81 and M74, can be explained by the density wave theory. However, it is difficult for the density wave theory to explain clearly for the flocculent spiral galaxies, such as M33 and NGC 7793, which have no obvious arms accumulation areas, but still show the spiral shapes(\citet{pue1992,kenn1981,sellwood2011}). For the initial phase of the particles is random, The simulating results according to the density wave theory should show a variety of non-spiral structures, but the ratio of the spiral galaxies is so high. To illustrate this case, the density wave theory is not sufficient. In addition to the above problems, there are still no reasonable explanations based on the density wave theory for the spiral galaxy with multi-arms structure and for the ubiquitous warped structure of the galaxy(\citet{terquem2000,kim2014,elmeg1982,grosb2004}). There are other theories, which do not yet explain the spiral structure very well(\citet{toomre1972,souza2013,sellwood1984,baba2013,reshe2016,gama1995}). 

The density wave theory is proposed mainly for the formation of the spiral arm. However, the existence of the spiral arm does not mean that the trajectory of the single moving particle will exhibit a spiral structure. It is considered in density wave theory that the particle in the spiral galaxies is doing the elliptical motion in the central potential. Nevertheless, even if the spiral structure is formed, it is difficult for the galaxy  to keep the stability and sustainability of the spiral arm. Relatively, a spiral galaxy tends to be stable. Both the local damage and the deformation of spiral structure are difficult to be observed(\citet{dobbs2014}). However, the observations show that the spiral structure of the galaxy is very stable, and no damage phenomenon of the structure is observed. Therefore, there must be other reasons for the formation of the spiral structure and the spiral arm.

In this work, a new view for the general existence of the spiral galaxies is proposed that it is due to the large scale rotation of the background, and then the spiral galaxy is not a relatively isolated system during its evolution process. The rotation of its physical background makes the galaxies more like  a cosmic hurricane.

Based on this view point, a physical mechanism which is very different from the density wave theory is proposed for the formation of the spiral structure in the spiral galaxy. the problem of the galactic evolution is reduced to a viscous two-body problem in a non-inertial system. The viscous two-body system is simulated for researching  the force upon the galactic particle and its influence to the galaxies. The trajectory of a single galactic particle is calculated under the actions of the gravitational force, the galaxy viscous force and the Coriolis force. The results show that the trajectory of galactic particle under the background rotation is a spiral structure itself, not an elliptic orbit. And the origin of both the spiral arm and the warping structure can be demonstrated self-consistently. Under the effect of the rotation background, the transverse velocity of the galactic particle rotating around the centre is mainly the result of the Coriolis force, rather than the effect of the rotating velocity of the elliptical orbit under the action of central gravitational force. And this transverse velocity does not need to be explained with dark mater(\citet{turner2000}). 

\section{The physical mechanism in the motion of Galaxies}\label{sect:mechanism}
In the density wave theory, the formation of spiral arm is a combined results of particles doing elliptical motion under the central potential, which actually considers the galaxies as a relatively isolated system. Aiming at this point, it is  proposed that the spiral galaxy in cosmos is not a relatively isolated statistical system, and its evolution is also driven with the motion of the background. A non-inertial force evoked from the universal background motion participates in the evolution of the galaxies, which is the significant reason for the galactic spiral structure. Therefore, the inertial force should be taken into account when the structure and evolution of the galaxies are simulated.
A galactic particle in its movement is attracted by other particles with the force of gravitation. Because of the central symmetry, the net force on every particle is centripetal. However, there is no enough huge mass in the center to provide this kind of gravitational force to make the particle moving in an elliptical orbit. Therefore, the galaxy is considered as a many-body system in classical mechanics while analyzing its kinetic mechanism.

a net viscous force is also provided from the interactions between the galactic particles for a single particle other than both the central net force and the non-inertial force exerted by the rotational background(\citet{lynden1974,hofner1994,new1998,flebbe1994}). This viscous force is actually produced from the gravitational forces of other matters around the particle, i.e., the viscous force is one component of the gravitational force. Up to now, there are no sufficient theoretical and experimental data to give the explicit expressions for the viscous force of the galaxies. Therefore, the viscous coefficient is taken as a free parameter, and the viscous force is considered as proportional to the velocity of the particle. The viscous force here is one kind of the inner gravitational force of the whole galaxies, so the total mechanical energy of the galactic system does not changed while a particle is under the viscous force. And the lost energy of the single particle will make the change of the rotational velocity of the whole galaxy. It is neglected in simulation cause its effect on the single particle is very little.

By doing so, the evolution of galactic system can be considered as a many-body problem under these three forces now. The spiral structure of the galaxy can be given by many multi-body and fluid simulations coming from some galactic theory such as the density wave theory. Nevertheless, the influence of an inertia force can not be clearly investigated by the more complicated multi-body and fluid simulations, and the relationship between the forces and the galactic evolution can not be comprehended intuitively. In this work, the kinetic properties of galaxy under those forces is analyzed with the simplest galactic system only consisting of two particles. The research on the evolution of the overall galaxy in detail requires further simulations of a system with more particles. 
However, the simplified two-particle galaxy can not only give the explicit conclusion, but also make the relationship of the forces and the evolution of the galaxy comprehended more intuitively and easier when the effect of those forces is investigated. Then the reduced galactic system consisted of two particles makes the conclusion more convincing. Therefore, it is necessary to carry out such analysis before the many-body or Hydrodynamic simulations is given.

The main forces exerted on the galactic particle are the gravitational force, viscous force and non-inertial force, so the net force on one galactic particle can be represented as
\begin{equation}            \label{eq1}
	\vec{F}={{\vec{F}}_{g}}+{{\vec{F}}_{v}}+{{\vec{F}}_{c}}
\end{equation}
Where ${{\vec{F}}_{g}}$ is the net gravitational force on the particle, ${{\vec{F}}_{v}}$ is the viscous force, and ${{\vec{F}}_{c}}$ is the non-inertial force exerted by the rotational physical background. The force ${{\vec{F}}_{c}}$ is an important factor for the galactic evolution here.
Because of the rotational symmetry of a spiral galaxy, the net gravitational force exerted on a galactic particle by all the other galactic particles is a centripetal force. In this simplified two-particle galaxy, the net gravitational force is represented with the gravitational force between two particles, so the gravitational term in equation (\ref{eq1}) can be written as
\begin{equation}            \label{eq2}
	{{\vec{F}}_{g}}=-G\frac{{{m}_{1}}{{m}_{2}}}{{{r}^{3}}}\vec{r}
\end{equation}
where $G$ is the gravitational constant,${{m}_{1}}$and${{m}_{2}}$ are the masses of the two particles respectively, $\vec{r}$ is the distance vector between two particles.

When a particle moves in a galaxy, it is also pulled by a residual gravitational force from other particles, which forms the viscous force exerted on this particle. The viscous force is introduced in our double particles galaxy model, and the viscous force is supposed to be proportional to the velocity of the particle empirically, i.e.
\begin{equation}            \label{eq3}
	{{\vec{F}}_{v}}=-\eta\vec{v}
\end{equation}
where $\eta$ is the viscosity coefficient for the galactic particles. According to the data from current galaxy researches, there is not yet a convincing numerical value for $\eta$, which is taken as a free parameter here.

As an assumption, it is not suitable to generate the inertial force from the cosmic background in a complicated way. There are two simple ways to provide an inertia force, one is the uniformly accelerated motion, and another is the uniform rotation. The same inertia acceleration is provided to each particle in uniformly accelerated motion, which can not exert an influence on the self evolution of the galaxy. Therefore, the influence of the background in uniformly accelerated motion is regardless. And the inertia force provided by the cosmic background in an uniform rotation is taken into account for the effect of the galaxy evolution. With respect to the large scale background rotation, the galactic scale $D$ is far less than the radius of the background rotation $R$, i.e., $D\ll R$. So the difference of centripetal force produced by the rotational background on each particle is omitted. Thus the inertial force on each galactic particle produced by the uniform rotation of the background is only Coriolis force, which can be expressed as 
\begin{equation}            \label{eq4}
	{{\vec{F}}_{c}}=-2m\vec{\omega }\times \vec{v}
\end{equation}
where $m$ is the mass of the galactic particle, and $\vec{v}$ is the velocity of galactic particle relative to the galactic centroid in the background rotation frame, and $\vec{\omega }$ is the angular velocity of the background rotation. 
In this case, the forces on the galactic particle under the large scale rotation of the background are one gravitational force, one viscous force and one Coriolis force. The net force can be expressed as
\begin{equation}            \label{eq5}
	\vec{F}=-G\frac{{{m}_{1}}{{m}_{2}}}{{{r}^{3}}}\vec{r}-\eta\vec{v}-2{{m}_{1}}\vec{\omega }\times \vec{v}
\end{equation}
Based on these reasonable assumption, a reduced galactic system consisted of two particles is produced. Those particles have the same mass because of the rotational symmetry of the galaxy, and the initial relative velocity of each particle is set as zero. The movement of the galactic particle can be numerically simulated with the net force in equation (\ref{eq5}).

\section{Simulation and analysis about viscous two-body problem}\label{simulation}
Because the main physical properties of the model is focused on here, the dimensionless unit is used in simulating the equation (\ref{eq5}), i.e., $G={{m}_{1}}={{m}_{2}}=m=1$. The dynamical equations for particles of a two-particle galaxy can be represented as
\begin{equation} \label{eq6}
\left\{ \begin{aligned}
  & \frac{{{d}^{2}}{{{\vec{r}}}_{1}}}{d{{t}^{2}}}=-\frac{{{{\vec{r}}}_{12}}}{r_{12}^{3}}-\eta{{{\vec{v}}}_{1}}-2\vec{\omega }\times {{{\vec{v}}}_{1}} \\ 
 & \frac{{{d}^{2}}{{{\vec{r}}}_{2}}}{d{{t}^{2}}}=\frac{{{{\vec{r}}}_{12}}}{r_{12}^{3}}-\eta{{{\vec{v}}}_{2}}-2\vec{\omega }\times {{{\vec{v}}}_{2}} \\ 
\end{aligned} \right.
\end{equation}
Where ${{\vec{r}}_{12}}$ is the displacement vector of particle 1 relative to particle 2, both ${{\vec{v}}_{1}}$ and ${{\vec{v}}_{2}}$ are the relative velocities of particle 1 and particle 2 in the non-inertial system, respectively, $\eta$ and $\vec{\omega }$ are the adjustable parameters. In the simulation, the trajectories of particles in equation (\ref{eq6}) can be calculated by setting the initial position and the adjustable parameters, and the properties of galactic evolution are investigated. 
The important results are given via the simulating calculation.The spiral structure is conspicuous and stable in the trajectories of particles. but also that the formation of the spiral arm can be explained specifically through relation between the velocity and the radius. Moreover, the unusual physical phenomenon, such as distinct warped structure, is given by the three-dimensional simulation, which is consistent with the real warped galaxies.

\subsection{spiral structure}\label{spiral}
In the case of plane motion, the direction of $\vec{\omega }$ is set as Z axis, and the centroid location of the two particles is taken as the original point. The line connecting two particles is perpendicular to $\vec{\omega }$, and the initial velocities of the particles are all zero, and then the motion of the system can be investigated.

\begin{figure*}[htb]
\centering
\begin{minipage}{18cm}
\centering
\includegraphics[scale=0.5,trim={17.cm 2cm 0cm 0cm},clip]{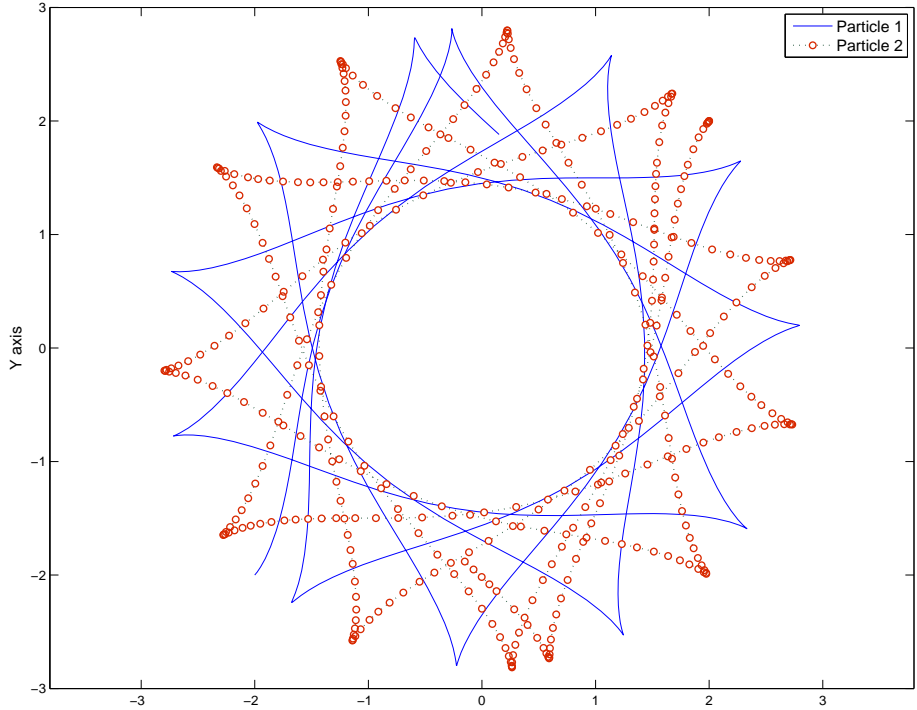}
\end{minipage}
\caption{Moving trajectories of double particles with $\eta=0$}\label{fig:nospiral}
\end{figure*}

By setting without the viscous force, i.e., $\eta=0$, and giving some initial conditions, there have no spiral structure in the trajectory calculated from the equation (\ref{eq6}), and the result is shown in figure \ref{fig:nospiral}. The trajectories show a petal-shaped structure with sharp end, which is similar to the trajectory of Foucault Pendulum with a zero initial velocity(\citet{some1972}). The reason for appearing this petal shape is that only the direction of the particle velocity is changed by Coriolis force, while the energy of the system can not be influenced. When the particle is attractive by another one and moves towards the center, the magnitude of the particle's velocity is increased, and the direction of the particle's velocity also keeps changing under the action of Coriolis force. 

At the position where is the nearest neighborhood from the center, the particle's radial velocity decreases to zero, and its tangential velocity reaches the maximum value. At this time, Coriolis force also reaches it's maximum value. At the tip position, the potential energy comes up to the maximum value, both the velocity and Coriolis force are zero. this process is repeated and the particle's moving direction is always changed by the Coriolis force. Then a petal-shaped structure around the center is formed in the trajectory of the particle. It can also be seen from  this trajectory that Coriolis force does not change the energy of the system.
when the viscosity coefficient does not equal zero, the galactic particle surrounds and tends to the original point because the particle's energy loses  continuously for the action of viscous force. Then a spiral structure appears in the trajectory of the particle, which is displayed in figure \ref{fig:nospiral}. 
\begin{figure*}[htb]
	\centering
	\begin{minipage}{18cm}
		\centering
		\includegraphics[scale=0.27,trim={3.5cm 1.5cm 0 2cm},clip]{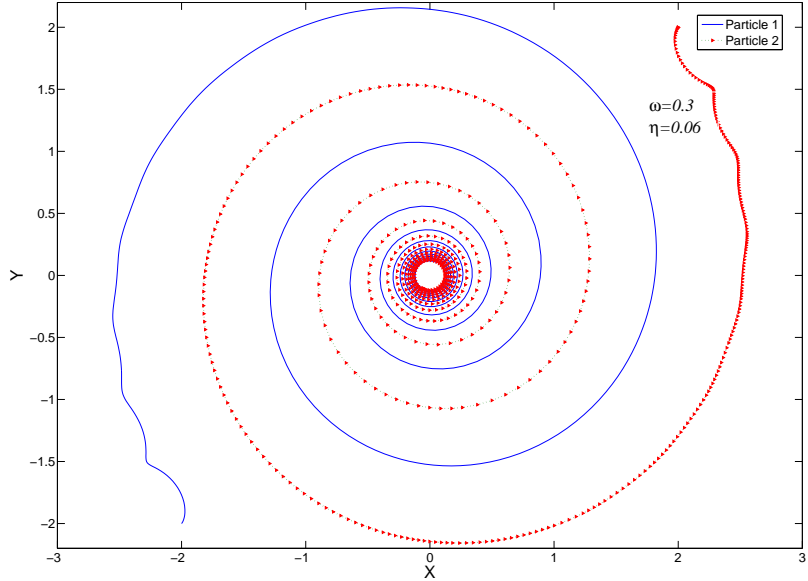}\vspace{0pt}
	\end{minipage}
	\caption{Trajectories of the double particles with $\eta=0.06$}\label{fig:spiral}
\end{figure*}

\begin{figure*}[htb]
\centering     
\subfigure[Hurricane Alberto]{\label{fig:hurr}\includegraphics[height=50mm]{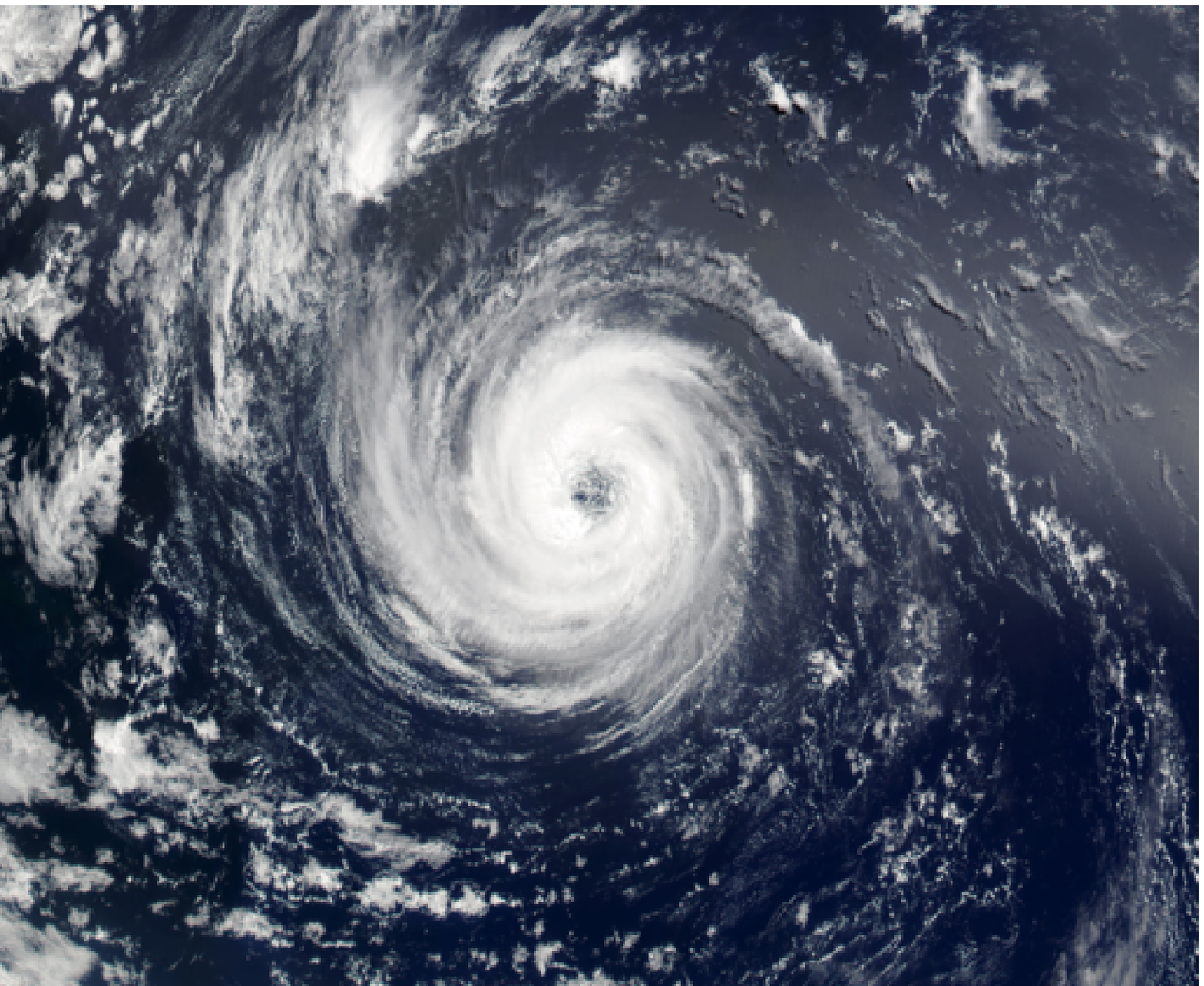}}
\hspace{6\medskipamount}
\subfigure[NGC 6814]{\label{fig:ngc6814}\includegraphics[height=50mm]{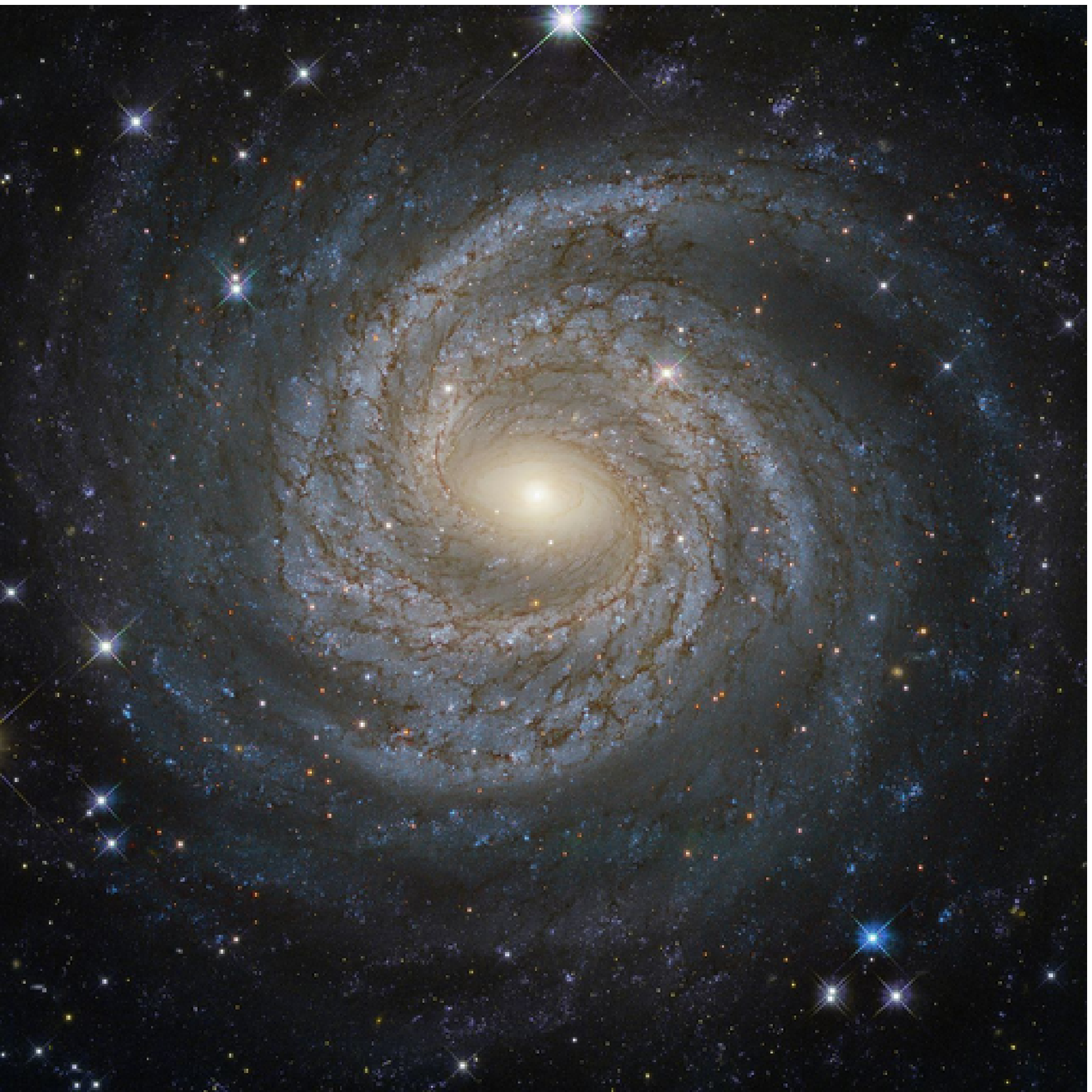}}
\caption{The similar spiral structure in Hurricane Alberto and NGC 6814}
\label{fig:hurricane}
\end{figure*}

Various shapes of the spiral structure can be given by adjusting the viscosity coefficient and the angular velocity of the galactic background, and these spiral structures are very similar to that in the spiral galaxies observed at present and the tropical hurricanes. The trajectory of the spiral structure in figure \ref{fig:spiral}, for instance, is extremely consistent with that of the spiral galaxy NGC 6814(or the Milk Way) and hurricane Alberto(2000) shown in figure \ref{fig:hurricane}(\citet{alberto,ngc6814}).

It is also showed that this spiral trajectory is very in accord with that of a logarithmic spiral function in the simulation calculation. One particle's trajectory in the double-particle galaxy with the initial position of $\text{(}\pm 5,\mp 5)$ is represented as that shown in figure \label{fig:log}. The trajectory can be fitted very well with the logarithmic spiral function $$r=2.015{{e}^{0.531\theta }}$$  
Therefore, that the initially static particle looks likes rotating around the center of galaxy, is the result of Coriolis force, rather than of be attracted by the galactic matter as an isolated system. Moreover, the previous assumption of introducing dark matter to explain the particle's motion as the effect of center gravitational force should be reevaluated. 

It is shown that the trajectory of the particle generated from  the equation (\ref{eq6}) has a spiral structure consistent with that of a spiral galaxy in the above analysis. Unlike in the density wave theory, the particle's trajectory is not an elliptical orbit, but is initially a spiral shape, which could explain why there exists a similar spiral structure in the flocculent galaxy without any explicit spiral arms. 
The spiral structure of the galaxies is so stable without any broken case, primarily because the trajectory of the particle is originally spiral, not an elliptical shape.
\begin{figure*}[htb]
\centering
\begin{minipage}{18cm}
\centering
\includegraphics[scale=0.35,trim={7.5cm 4.5cm 8 2cm},clip]{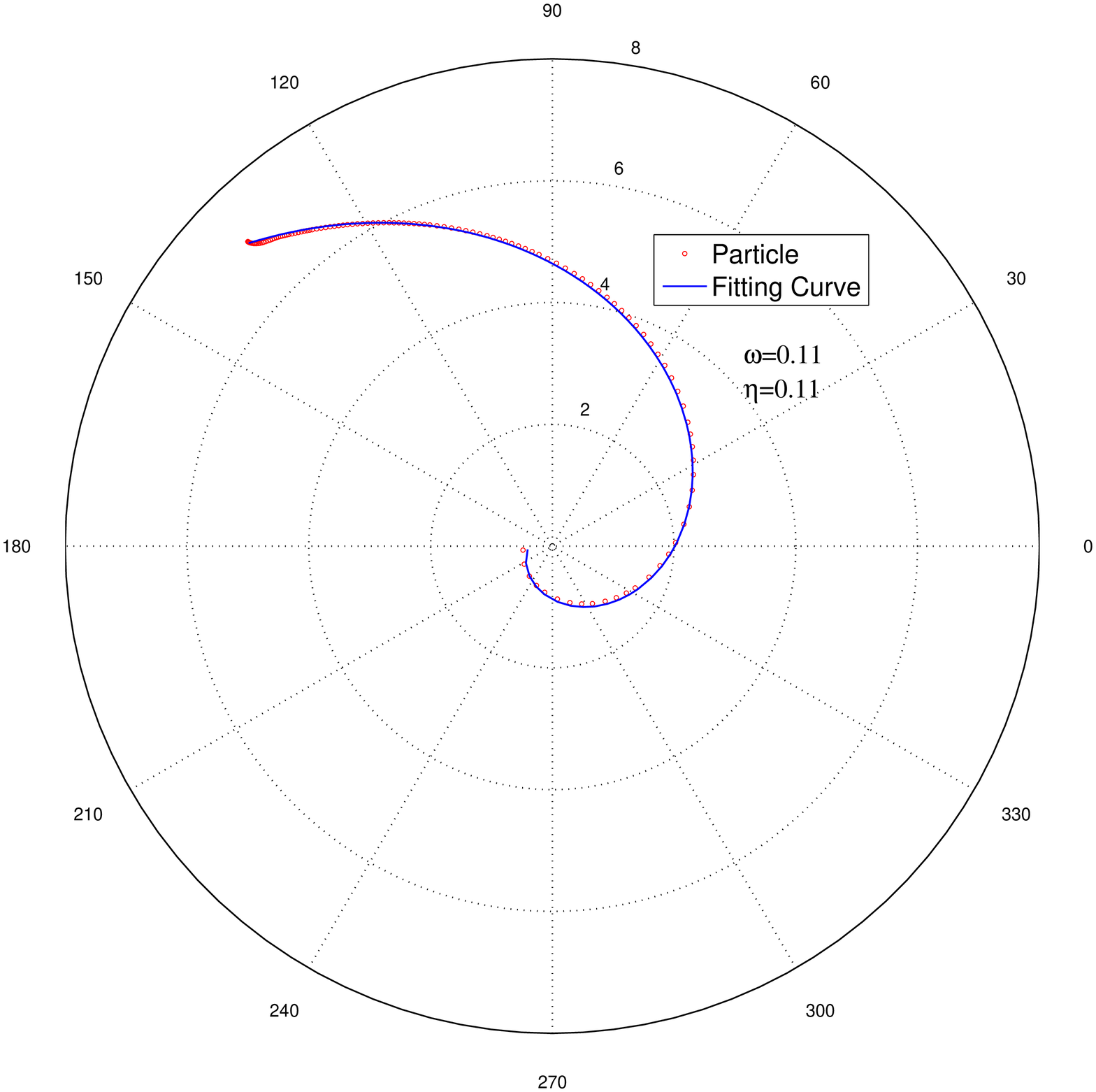}\vspace{0pt}
\end{minipage}
\caption{The logarithmic spiral curve fitting for a particle's trajectory}\label{fig:log}
\end{figure*}
\subsection{The formation and evolution of spiral arm}\label{spiralarm}
\begin{figure*}[htb]
\centering
\begin{minipage}{18cm}
\centering
\includegraphics[scale=0.3,trim={1.5cm 2cm 0 2cm},clip]{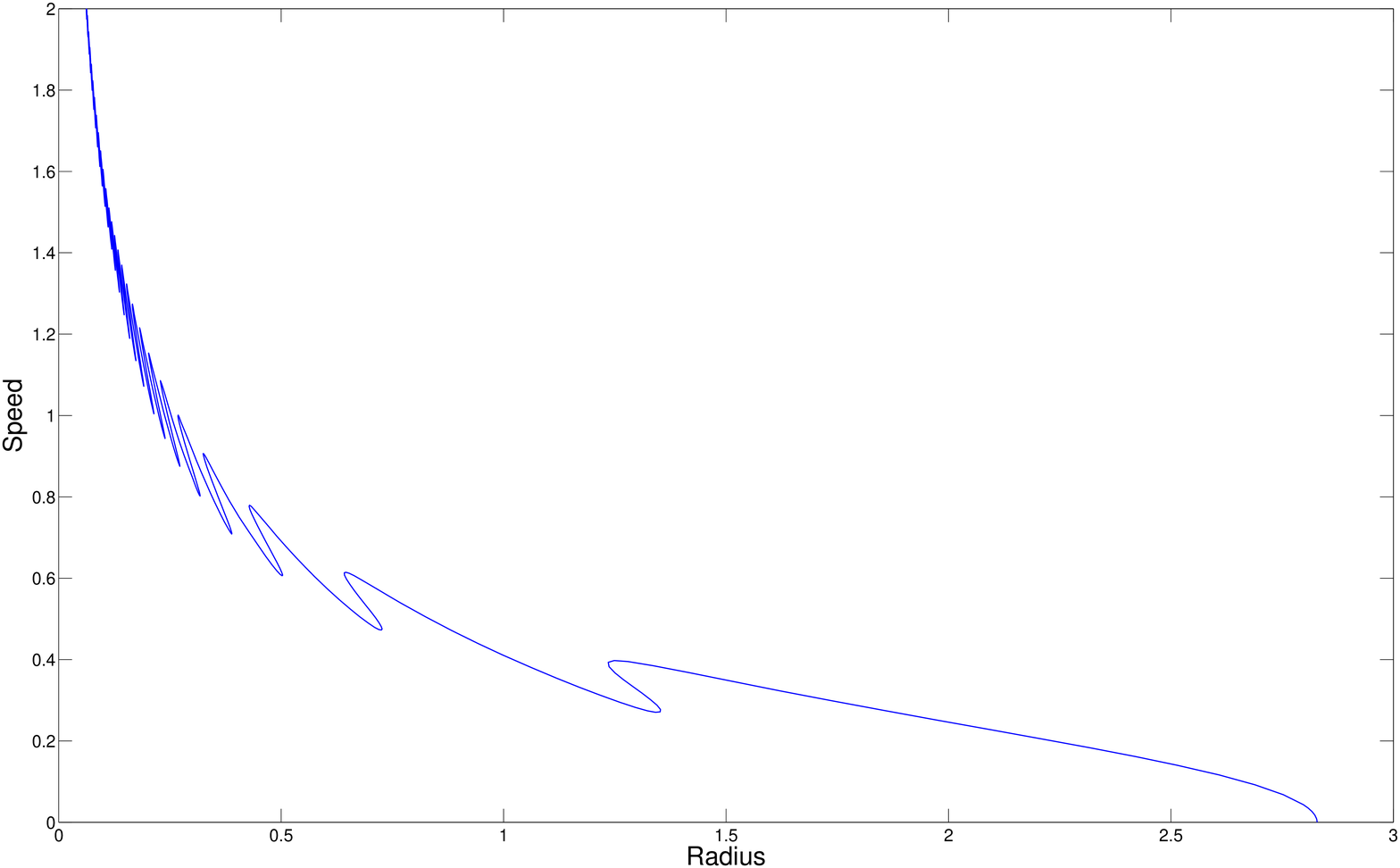}\vspace{0pt}
\end{minipage}
\caption{the relationship between velocity and radius of particle motion}\label{fig:velocity}
\end{figure*}

It is impossible to simulate the formation of the spiral arm in a galaxy of two particles, because the number of particles is too small. Nevertheless, the velocity changing with the radius is represented in figure \ref{fig:velocity}, and it can be seen that the velocity-radius curve of the particle have some turning-back. Considering the trajectories with $\eta=0$ in figure \ref{fig:nospiral}, it can be observed that the turning-back area actually change from those endpoint driven by both the Coriolis force and the viscous force. 
While the particles in a galactic disc move along the trajectories of the spiral curve, the foregoing particle with the turning-back will gather together near the orbit with the subsequent particle in the same trajectory and those nearby. As a result, the higher density area is generated, which then forms the galactic spiral arm.
The hurricane on the earth also has the similar spiral shape and the similar spiral arm under the action of Coriolis force generated from the earth spin. Just because of this turning-back of the velocity, the clouds area with different density is generated and forms the spiral arm-like structure of the hurricane. Furthermore, the hurricane's spiral arm can be explained only with Coriolis force, while both the density wave theory and the dark matter are needless.

If the large-scale background rotation existed, the spiral arm of the galaxy should be formed by the turning-back driven by Coriolis force, and which is very different from the explanation of the density wave theory. Moreover, all the particles move steadily along their spiral trajectories, which could explain the general existence of the spiral arm and the time stability of the spiral structure. the broken spiral arm caused by the moving particle in an elliptical orbit supposed in density wave theory has never been observed in astronomy. 

Further, the turning-back drives the formation of the spiral arm, and along the circumference there have many turning-back location which can be seen in figure \ref{fig:nospiral}, so the arm number in the original galaxy might be more than two.
However, during the evolution of the galaxy's multi-arm structure, the nearby spiral arms get together continuously due to the sustained impact of those turning-back, and the number of arms is reduced to two. 
\subsection{The origin of thin disc structure}\label{disc}
The thin disc structure is an significant feature of the spiral galaxies. By analysis the physical model in equation \ref{eq6}, it is obtained that the formation of the disc structure is directly associated to the viscous force $-\eta \vec{v}$ and Coriolis force $\vec{\omega }\times \vec{v}$.
During the particles falling to the center, the velocity component perpendicular to $\vec{\omega }$ reduces slowly due to the turning-back evoked by Coriolis forces. However, in the direction parallel to $\vec{\omega }$, there have no turning back for the component of Coriolis force being zero, and this energy loss is more than the one perpendicular to $\vec{\omega }$ under the only viscous force at the same time. The velocity component parallel to $\vec{\omega }$ decrease more rapidly.

In this case, the particles in the direction parallel to the $\vec{\omega }$ accumulate in the thin disc plane soon. however, the particles perpendicular to the $\vec{\omega }$ direction are still doing spiral motion far from the center at the same time. Therefore, a thin disc is formed before long during the evolution of the galaxy by this way.

\subsection{The formation of the warped structure}\label{warped}

Through setting the direction of angular velocity of the background rotation as the z-axis, and making the straight line between the two particles no longer in the same plane perpendicular to the $Z$-axis, the 3D trajectory of particles is simulated too.
In addition to the case of plane motion, the three-dimensional simulation also give some important conclusion, especially the warped structure which usually appears in the trajectories of the two-particle galaxy. Therefore, the 3D trajectory is no longer a kind of plane distribution.

\begin{figure*}[htb]
\centering
\begin{minipage}{18cm}
\centering
\includegraphics[scale=0.35,trim={4.cm 1.5cm 0 2cm},clip]{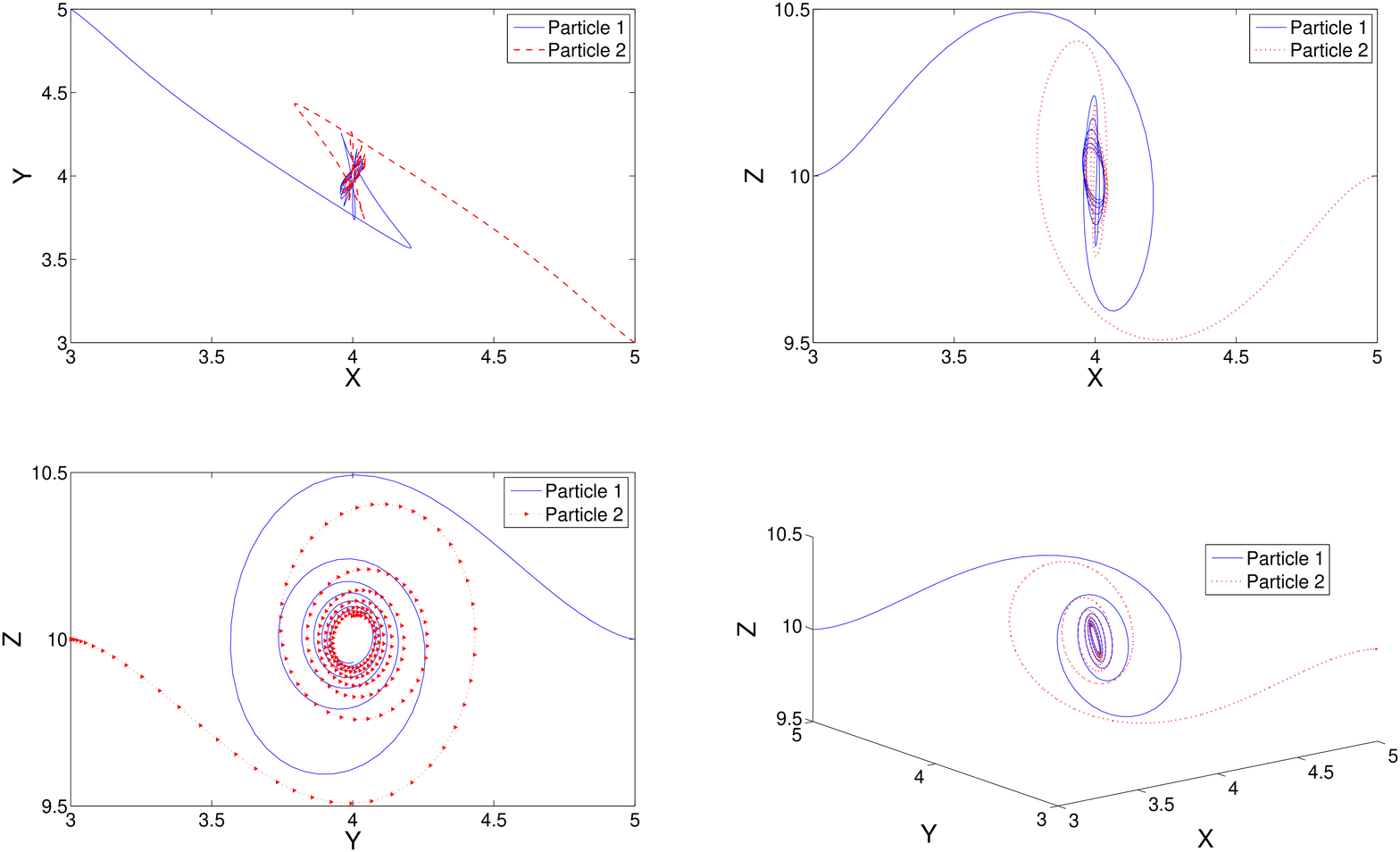}
\end{minipage}
\caption{3D trajectory and its projections of two particles' motion}\label{fig:traj}
\end{figure*}

The three-dimensional simulation of the equation (\ref{eq6}) can be proceeded by the following setting: the initial positions of the two particles are located at ${{\vec{r}}_{10}}=(3,5,10)$ and ${{\vec{r}}_{20}}=(5,3,10)$, respectively; the initial velocities are all zero; and the angular velocity of the background rotation is set as \[\vec{\omega }=\left( 0.36\cos (\frac{\pi }{6}),0,0.36\sin (\frac{\pi }{6}) \right)\] which is no longer set as z-axis; the viscosity coefficient is set as $\eta =0.16$.
The three-dimensional spatial trajectory of the particles is shown in figure \ref{fig:traj}. It can be seen distinctly from the $XY$ projection that a spatial warped structure is in the particles' 3D trajectory. And the different shapes of the warped structure can be also simulated in the galactic particles' trajectories by setting some different initial parameters.

When the angular velocity of the background rotation is not parallel to the normal of the rotational galactic disc, there is a component ${{\vec{\omega }}_{\bot }}$ parallel to the rotational disc, which occurs a component of Coriolis force $m{{\vec{\omega }}_{\bot }}\times \vec{v}$ perpendicular to the rotational plane. This force component is called as the leaving-plane force here. The particle will move away from the rotational plane along its normal direction under the action of this leaving-plane force. 

The leaving-plane force reaches the maximum value when the particle's velocity are vertical to  ${{\vec{\omega }}_{\bot }}$, and it is zero when the particle's velocity is parallel to ${{\vec{\omega }}_{\bot }}$. 
It can be seen that the leaving-plane force is anti-centrosymmetric in the disc. The direction of the leaving-plane force on one point is opposite to the one on its centrosymmetric point. And the leaving-plane force is changed with the particle's velocity. In this way, the warped structure is formed by the action of this component of the Coriolis force.

It is always a tough task to explain the warped structure of the galaxies, and most of them need to employ dark matter (\citet{lopez2002, deba1999, kerr1957, binney1978}). Different from the current theory on the warped structure, the spatial warped structure is derived  naturally from the background rotation in this work, and no additional assumption is required. The influence of the related parameters such as the background rotational frequency on the warped structure is also investigated through the numerical simulations.
It is shown in the simulating results that the warped structure is a general phenomenon except the normal direction of the disc completely parallel to the angular velocity of the background rotation. This conclusion is consistent with the astronomical observations(\citet{gama1995,sanchez1990}).

Due to the existence of the real galactic disc, it should be noted that the warped structures is different noticeably between the two-particle galaxy and the many-particle galaxies. The more results approach to the real galaxies need to be investigated through the N-body simulation. However, on the physical relationship between Coriolis force and the warped structure, the result given in this work is more intuitive. If the background were rotating, the warped structure should appear certainly in the real galaxies.
\subsection{The whole turning of the galactic disc}\label{turning}
In addition to the formation of the warped structure, the whole turning of the thin galactic disc can be evoked by  a torque $\vec{r}\times (m\vec{\omega}_{\bot } \times \vec{v})$ produced by the leaving-plane force ${{\vec{\omega }}_{\bot }}$. 
The sum torque vector is in the thin disc plane and passes through the center of the thin disc, and is perpendicular to ${{\vec{\omega }}_{\bot }}$. The direction of the whole turning of the disc, should be parellel to the torque vector.

It might be an important evidence of the overturning of the disc that the jet-current from the center of Centaurus A turns into a form with the s-shape through processional motion. And if there exists the background rotation, the circumferential band formed by the historical rudiments, i.e., a small amount of the material left during the galactic overturn process, and the other related phenomenons about the overturning could be observed as usual.  Those reference need more observed evidence to prove. 

%
\section{Conclusions and prospects}\label{conclusion}
A model of the simplified galaxy for its evolution is simulated and analyzed with the existence of the background rotation in this work. The research of the whole galactic evolution still needs the further many-body simulation or the  hydrodynamic  simulation. However, with regard to the properties of the forces and the resulting spiral structures, spiral arms and warped structures and so on, the effects of forces can be demonstrated more clearly here, and the conclusions  are more convincing for the simplicity and intuition of the model.

In the density wave theory, the particle is doing an elliptical motion, and the particle's velocity is periodic. The spiral arm is made up of areas of greater density. Furthermore, the assumption of dark matter is introduced for explaining the huge centripetal gravitational force for the rotation. However in this work, the trajectory of a single particle is itself spiral, and the high tangential velocity is associated to Coriolis force and viscous force, not contributed by the huge centripetal gravitational force. Therefore, the dark matter is not required yet. If there is a large scale rotation of the galactic background indeed, the assumption of  dark matter needs to be reevaluated.

In addition, different from the explanation of the spiral arm in density wave theory,  the spiral arm is associated to turning-back of the particle's velocity in this work.  the  turning-back  makes the particles in the same spiral trajectory and the nearby areas  aggregate, and impulses the form of spiral arm. Moreover, this model can also be applied to explain the spiral arm structures of the tropical hurricane on the earth, which can not be demonstrated by the density wave theory under the centripetal gravitational force obviously.

A tropical hurricane on the earth is mainly a two-dimensional structure as it keeps close to the ground. However, the three-dimensional galaxies in the background rotation also have some physical phenomena worthy of attention, such as  warped spiral structure, spiral arm and the whole turning of the disc(\citet{burns1983}). Although the background rotation need to be  confirmed by further astronomical observations, if the background rotation exists, it can be supposed sufficiently that the spiral galaxy is exactly a cosmic hurricane under the background rotation. According to the widespread distribution of the spiral galaxies, the large-scale rotation in the universe can basically be regarded as a normal and general phenomenon.

In our previous work(\citet{cao2003}), the horizon problem of the rotation velocity  within the Galileo rotating framework had been solved in the special relativistic rotating frame. No horizon problem restricts a large-scale rotation in the cosmos now. The results in that paper showed that the large-scale rotation of the galactic background can exist. Therefore, the percentage that the transverse Doppler red shift formed by the background rotation occupies in the Hubble red shift is also a problem worthy of discussing.
\acknowledgements
We appreciate the helpful comments from the anonymous referee. We thank the students CAI Yuanqiang, GONG Jiguang, GUO Juanjuan and ZHANG Jingyu for their helpful work in computer, and we also thank WANG Heng, JU Liping for their helpful discussions and comments.

\end{document}